\documentclass[twocolumn,showpacs,preprintnumbers,amsmath,amssymb]{revtex4}
\begin{document}

\title{Action principle for Numerical Relativity evolution systems}

\author{C.~Bona$^{1}$, C.~Bona-Casas$^{1}$ and C.~Palenzuela$^{2,3}$}
\affiliation{$^1$Institute for Applied Computing with Community
Code (IAC$^{\,3}$).\\
Universitat de les Illes Balears, Palma de Mallorca 07122, Spain
\\
$^2$Canadian Institute for Theoretical Astrophysics (CITA),
Toronto, Ontario M5S 3H8, Canada.
\\
$^3$Dept. of Physics \& Astronomy, Louisiana State University,
Baton Rouge, LA 70803, USA. }


\begin{abstract}
A Lagrangian density is provided, that allows to recover the Z4
evolution system from an action principle. The resulting system is
then strongly hyperbolic when supplemented by gauge conditions
like '1+log' or 'freezing shift', suitable for numerical
evolution. The physical constraint $Z_\mu = 0$ can be imposed just
on the initial data. The corresponding canonical equations are
also provided. This opens the door to analogous results for other
numerical-relativity formalisms, like BSSN, that can be derived
from Z4 by a symmetry-breaking procedure. The harmonic formulation
can be easily recovered by a slight modification of the procedure.
This provides a mechanism for deriving both the field evolution
equations and the gauge conditions from the action principle, with
a view on using symplectic integrators for a constraint-preserving
numerical evolution. The gauge sources corresponding to the
'puncture gauge' conditions are identified in this context.
\end{abstract}

\maketitle

\section{Introduction}
The role of action principles is so crucial in theoretical physics
that its importance can not be overemphasized. In the case of the
General Relativity, the standard action was proposed by Hilbert
since the very beginning of the theory, although the Hamiltonian
formulation had to wait for decades~\cite{Dirac58,ADM}. The reason
for this long delay is probably related to the complexity of the
Cauchy problem for Einstein's equations, which becomes manifest in
the 3+1 (space plus time) decomposition~\cite{ADM}. The coordinate
gauge freedom produces a mismatch between the number of dynamical
fields and that of true evolution equations: four of the field
equations are indeed (energy-momentum) constraints. This rich
structure opens the door to many different approaches.

On the other hand, by the end of the past century, some hyperbolic
extensions of Einstein's equations were developed with a view on
numerical relativity applications~\cite{CR83,BM92,BM95,AACY95}.
This emergent field is now more mature: there are two main
formalisms currently used in numerical simulations. One is
BSSN~\cite{SN95,BS99}, working at the 3+1 level, and the other is
the class of generalized harmonic
formalisms~\cite{Frie85,Pret05,Lind06}, working at the
four-dimensional level. A unifying framework is provided by the Z4
formalism~\cite{Z4}, which allows to recover the generalized
harmonic one by relating the additional vector field $Z_\mu$ with
the harmonic '\,gauge sources'~\cite{Frie85}. On the other hand,
it allows to recover (a specific version of) BSSN by a
symmetry-breaking process in the transition from the
four-dimensional to the three-dimensional
formulations~\cite{Z48,LNP2}.

There is a growing interest in incorporating the new hyperbolic
formulations into the Lagrangian/Hamiltonian framework. An example
is the usage of the '\,densitized lapse'~\cite{CR83} as a
canonical variable, leading to a modification in the standard form
of the canonical evolution equations~\cite{AY98}. Reciprocally,
there are very recent attempts of modifying the ADM action in
order to incorporate coordinate conditions of the type used in
numerical relativity~\cite{B08,HR10}, with a view on using
symplectic integrators for the time evolution, which could ensure
constraint preservation in numerical simulations~\cite{BGP05}. On
a different context, a well posed evolution formalism developed
from a Lagrangian formulation could be a good starting point for
Quantum Gravity applications.

In this paper we derive for the first time the Z4 formalism from
an action principle by introducing a Lagrangian density which
generalizes the Einstein-Hilbert one. We also provide the
corresponding Hamiltonian, via the Legendre transformation. This
is a crucial step towards the Hamiltonian formulation of other
numerical-relativity formalisms, like BSSN. On the other hand, we
recover the generalized harmonic formulations as usual, by
relating the additional vector field $Z_\mu$ with the harmonic
'\,gauge sources'. This mechanism is generalized, by identifying
the gauge sources which correspond to the current
numerical-relativity coordinate conditions, as we show explicitly
for the 'puncture gauge': the combination of the '\,1+log' slicing
and the gamma-driver conditions.

\section{The action principle}

Let us consider the generic action
\begin{equation}\label{action}
   S = \int d^4x~{\cal L}
\end{equation}
with a Lagrangian density which generalizes the Einstein-Hilbert
one by including an extra four-vector $Z_\mu$, namely
\begin{equation}\label{Lagrangian}
    {\cal L} = \sqrt{g}~g^{\mu\nu}\,[\,R_{\mu\nu}+2\,\nabla_\mu
    Z_\nu\,]
\end{equation}
(we restrict ourselves to the vacuum case), with the Ricci tensor
written in terms of the connection coefficients
\begin{equation}\label{Ricci}
    R_{\mu\nu} = \partial_{\rho} \Gamma^{\rho}_{\mu\nu}
      - \partial_{(\mu} \Gamma^{\rho}_{\nu)\rho}
      + \Gamma^{\rho}_{\rho\sigma}\, \Gamma^{\sigma}_{\mu\nu}
      - \Gamma^{\rho}_{\sigma\mu}\, \Gamma^{\sigma}_{\rho\nu} ~,
\end{equation}
(round brackets denote symmetrization).

Now let us follow the well-known Palatini approach, by considering
independent variations of the metric density
$h^{\mu\nu}=\sqrt{g}~g^{\mu\nu}$, the connection coefficients
$\Gamma^\rho_{\mu\nu}$ and the vector $Z_\mu$. From the
$h^{\mu\nu}$ variations we get directly the Z4 field
equations~\cite{Z4}
\begin{equation}\label{Z4}
    R_{\mu\nu}+\,\nabla_\mu Z_\nu +\,\nabla_\nu Z_\mu = 0\,,
\end{equation}
which are currently used in many numerical-relativity
developments~\cite{LNP2}.

From the $\Gamma^\rho_{\mu\nu}$ and the $Z_\mu$ variations we get
a coupled set of equations which can be decomposed in a covariant
way into the tensor equation
\begin{equation}\label{metric connection}
    \nabla_\rho\,g^{\mu\nu} = 0\,,
\end{equation}
which fixes the connection coefficients in terms of the metric,
and the vector condition
\begin{equation}\label{Zis0}
    Z_\mu =0.
\end{equation}
Let us note here the different role of the conditions (\ref{metric
connection}) and (\ref{Zis0}). As there are much more independent
connection coefficients than evolution equations in (\ref{Z4}), we
will consider condition (\ref{metric connection}) as a constraint
enforcing the metric connection '\,a posteriori', that is after
the variation process. In this way, we will ensure that equations
(\ref{Z4}) are identical to the original Z4 equations, rather than
some affine generalization. For this reason, we will assume a
metric connection everywhere in what follows.

The case of condition (\ref{Zis0}) is different, as the Z4
equations (\ref{Z4}) actually provide evolution equations for
every component of $Z_\mu$. Then, (\ref{Zis0}) is a standard
primary constraint and we have a choice among different strategies
for dealing with it. If we enforce (\ref{Zis0}) into the Z4 field
equations (\ref{Z4}), we get nothing but Einstein's equations.
This is not surprising because our Lagrangian obviously reduces to
the Einstein-Hilbert one when $Z_\mu$ vanishes. The problem is
that the plain Einstein field equations do not lead directly to a
well-posed initial data problem. This is why the original harmonic
formulation~\cite{DeDo21,Lanczos22,Fock59} was used instead in the
context of the Cauchy problem~\cite{Choquet52}. For the same
reason, other formulations (BSSN~\cite{SN95,BS99}, generalized
harmonic~\cite{Frie85,Pret05,Lind06}, Z4~\cite{Z4,Z48}) are
currently considered in numerical relativity.

We can alternatively follow a different strategy. Instead of
enforcing (\ref{Zis0}), we can deal with this condition as an
algebraic restriction to be imposed just on the initial data, that
is
\begin{equation}\label{Zis0weak}
     Z_\mu~|_{t=0} = 0
\end{equation}
In this way, we can keep the Z4 field equations which, when
supplemented with suitable coordinate conditions, lead to a
strongly hyperbolic evolution system~\cite{Z48,LNP2}. The
consistency of this 'relaxed' approach requires that the
constraint (\ref{Zis0}) should be actually preserved by the Z4
field equations (\ref{Z4}). In this way, the solutions obtained
from initial data verifying (\ref{Zis0}) will actually minimize
the proposed action (\ref{action}).

Allowing for the conservation of the Einstein tensor, which is
granted after the metric connection enforcement, we derive from
(\ref{Z4}) the second-order equation, linear-homogeneous in $Z$
\begin{equation}\label{Z4div}
    \nabla_\nu\,[\,\nabla^\mu Z^\nu +\nabla^\nu Z^\mu
    - (\nabla_\rho Z^\rho)\,g^{\mu\nu}\,]
    = 0\,.
\end{equation}
It follows that the necessary and sufficient condition for the
preservation of the constraint (\ref{Zis0}) is to impose also its
first time-derivative conditions in the initial data, that is
\begin{equation}\label{Emom}
(\partial_0\,Z_\mu)\,|_{t=0} = 0\,.
\end{equation}
Note that, allowing for (\ref{Zis0weak}) and the Z4 field
equations, the secondary constraints (\ref{Emom}) amount to the
standard energy and momentum constraints, which are then to be
imposed on the initial data in addition to (\ref{Zis0weak}).

This '\,relaxed' treatment of the constraints (\ref{Zis0}) may
look unnatural. But it is just the reflection of a common practice
numerical relativity ('\,free evolution' approach), where four of
the ten field equations (the energy-momentum constraints) are not
enforced during the evolution, being imposed just in the initial
data instead. The introduction of the extra four-vector in the Z4
formalism actually provides a simpler implementation of the same
idea.

\section{Hamiltonian formalism}

A detailed look at the Lagrangian density (\ref{Lagrangian}) shows
that the time derivatives of most of the variables are not present
in ${\cal L}$. The only exceptions are the combinations
\begin{equation}\label{time_deriv}
    \Gamma^0_{\,\mu\nu}-\delta^0_{(\mu}\Gamma^\rho_{\nu)\rho}
    +2\,\delta^0_{(\mu}\,Z_{\nu)}\,.
\end{equation}
This dynamical subset of variables can be decomposed into
\begin{equation}\label{canonical_vars}
   \{~\Gamma^0_{\,ij}\,,~ Z_i-\frac{1}{2}~(\Gamma^k_{ki}-\Gamma^0_{\,0i})
    \,,~ Z_0-\frac{1}{2}~\Gamma^k_{k0}~\}~~(i,j,k=1,2,3)\,,
\end{equation}
The corresponding canonical momenta are given, respectively, by
\begin{equation}\label{canonical_mom}
    \{~h^{ij}\,,~ 2\,h^{0i}\,,~ 2\,h^{00}~\}\,.
\end{equation}
The remaining quantities
\begin{equation}\label{multipliers}
    \{~\Gamma^k_{\,\mu\nu}\,,~\Gamma^0_{\,0\,\mu} \}
\end{equation}
can be considered as a sort of Lagrange multipliers, introducing
constraints into the dynamical system, as their time derivatives
do not enter the Lagrangian (see for instance
ref.~\cite{Lanczos_book}).

Note that our Lagrangian (\ref{Lagrangian}) is linear in the time
derivatives. This means that the relationship between field
velocities and momenta can not be inverted. The canonical momenta
corresponding to (\ref{canonical_mom}, \ref{multipliers}) vanish
identically, and the ones corresponding to (\ref{canonical_vars})
coincide with the metric density components (\ref{canonical_mom}).
This means that the Lagrangian (\ref{Lagrangian}) is a singular
one: the corresponding (constrained) canonical formalism can be
developed following the work of Dirac~\cite{Dirac64}.

We will rather sketch here a simpler approach, by performing a
limited Legendre transform, in the sense that it will only affect
the dynamical subset (\ref{canonical_vars}), with canonical
momenta (\ref{canonical_mom}). The remaining quantities will be
considered as Lagrange multipliers, for which no Legendre
transformation is required. We obtain in this way the Hamiltonian
function
\begin{eqnarray}\label{Hamiltonian}
    {\cal H} = -h^{\mu\nu}\{~\partial_k\,[~\Gamma^k_{\mu\nu}
    -\delta^k_\mu\,(\Gamma^\rho_{\rho\nu}-2Z_\nu)~] &&\\
   +  (\Gamma^\rho_{\rho\sigma}-2Z_\sigma)\Gamma^\sigma_{\mu\nu}
   - \Gamma^\rho_{\sigma\mu}\Gamma^\sigma_{\rho\nu}~\}\,,\nonumber &&
\end{eqnarray}
where the metric densities are considered here as the canonical
momenta associated to the dynamical fields (\ref{canonical_vars}).

The Hamilton equations for the fields (\ref{canonical_vars}) are
precisely the Z4 equations (\ref{Z4}). The Hamilton equations for
their momenta (\ref{canonical_mom}) can be written as
\begin{eqnarray}
\label{ham_eq a}
  \partial_\mu\,h^{\mu 0}&=& -\,h^{\mu \nu}~\Gamma^0_{\,\mu\nu} \\
\label{ham_eq b}
  \partial_\mu\,h^{\mu i}&=& -\,h^{\mu \nu}~\Gamma^i_{\,\mu\nu} \\
\label{ham_eq c}
  \partial_0\,h^{ij}&=& h^{ij}\,(\Gamma^\rho_{\,\rho 0}-2Z_0)
  -2\,h^{\rho(i}\,\Gamma^{j)}_{\,\rho 0}\,,
\end{eqnarray}
which must be supplemented with the constraints derived from the
Lagrange multipliers subset (\ref{multipliers}). A straightforward
calculation shows that the full set of Hamilton equations is still
equivalent to the Z4 equations (\ref{Z4}), plus the metric
connection requirement (\ref{metric connection}), plus the
vanishing of $Z_\mu$ (\ref{Zis0}). Indeed, allowing for
(\ref{metric connection}, \ref{Zis0}), the subsystem (\ref{ham_eq
a}-\ref{ham_eq c}) is verified identically.


Note that in all our developments we have preserved general
covariance. Our action integral (\ref{action}) is a true scalar
and, in spite of other alternatives, we have avoided the addition
of total divergences which could have simplified our developments
to some extent, at the price of adding boundary terms. This means
that we keep at this point the full coordinate-gauge freedom.

This is reassuring from the theoretical point of view, but it can
be a disadvantage if one is planning to use symplectic integrators
for numerical evolution, as the required coordinate conditions
must be supplied from the outside of the canonical formalism. This
is why some recent works are trying to incorporate the coordinate
conditions, via Lagrange multipliers, into the canonical
framework~\cite{B08,HR10}.

\section{Generalized Harmonic systems}

There is still another possibility, which allows a more direct
specification of a coordinate gauge at the price of breaking the
covariance of the evolution equations. which are currently used in
many numerical-relativity developments~\cite{LNP2}.
We can enforce in the Z4 equations (\ref{Z4}) the following assignment
for $Z_\mu$
\begin{equation}\label{harm}
    Z^\mu =
    -\frac{1}{2}~\Gamma^\mu_{\rho\sigma}\,g^{\rho\sigma}
    \equiv -\frac{1}{2}~{\Gamma^\mu}\,.
\end{equation}
The vanishing of $Z_\mu$ will amount in this way to the 'harmonic
coordinates' condition, which can be considered then as a
constraint to be imposed just in the initial data, that is
\begin{equation}\label{Harm_constraint}
    ({\Gamma^\mu}_{\rho\sigma}\,g^{\rho\sigma})\,|_{t=0} = 0
\end{equation}
(note that the extra field $Z_\mu$ has disappeared in the
process). The resulting field equations
\begin{equation}\label{Harmonic field eqs}
    R_{\mu\nu}-\,\partial_{(\mu} \Gamma_{\nu)}
    + \Gamma^{\rho}_{\mu\nu}\,\Gamma_{\rho} = 0
\end{equation}
lead, after imposing the metric connection condition (\ref{metric
connection}), to the manifestly hyperbolic second-order system
\begin{equation}\label{Z4_metric_connection}
    g^{\rho \sigma} \partial^2_{\rho\sigma}~g_{\mu\nu}
    =  2\,g^{\rho\sigma} g^{\alpha\beta}\, [
        \partial_{\alpha} g_{\rho \mu}~ \partial_{\beta} g_{\sigma \nu}
      - \Gamma_{\mu \rho \alpha} ~ \Gamma_{\nu \sigma \beta}]\,.
\end{equation}
This corresponds to the classical harmonic formulation of General
Relativity~\cite{DeDo21,Lanczos22,Fock59}, which is known to have
a well-posed Cauchy problem~\cite{Choquet52}.

We have derived in this way the harmonic formalism through the
non-covariant prescription (\ref{harm}). The harmonic constraint
(\ref{Harm_constraint}) is automatically preserved by the
resulting (harmonic) evolution system, provided that we also
enforce the energy-momentum constraints on the initial data. This
can be seen in a transparent way by replacing directly
(\ref{harm}) into the covariant constraint-evolution equation
(\ref{Z4div}) and then into the resulting conditions (\ref{Emom}).

The prescription (\ref{harm}) can be generalized in order to
enforce other coordinate gauges that are also currently used in
numerical relativity. The simpler
formulations~\cite{Frie85,Lind06} correspond to the replacement
\begin{equation}\label{harm_gs}
    Z^\mu =
    -\frac{1}{2}~(\,\Gamma^\mu + H^\mu\,)\,,
\end{equation}
where the '\,gauge sources' $H^\mu$ are explicit functions of the
metric and/or the spacetime coordinates. More general choices of
$H^{\mu}$, like that of ref.~\cite{Pret06}, would require a more
elaborate treatment.

The same mechanism can be applied to coordinate conditions derived
in the 3+1 framework, where the spacetime line element is
decomposed as
\begin{equation}\label{line3+1}
    ds^2 = - \alpha^2~dt^2
    + \gamma_{ij}~(dx^i+\beta^i~dt)~(dx^j+\beta^j~dt)\,.
\end{equation}
The spacetime slicing is given by the choice of the time
coordinate. In this context, the harmonic slicing condition can be
generalized to~\cite{Z48}
\begin{equation}\label{gen_harm}
(\partial_t -\beta^k\partial_k)~\alpha =
- f\, \alpha^2~tr K\,,
\end{equation}
where $K_{ij}=-\alpha\,\Gamma^0_{\,ij}~$ stands for the extrinsic
curvature of the time slices. The case $f=1$ corresponds to the
harmonic time-coordinate choice, whereas the choice $f=2/\alpha$
corresponds to the popular '\,1+log' time slicing.

In order to get the replacement, of the form (\ref{harm_gs}),
which connects this condition with our formulation, we must
rewrite (\ref{gen_harm}) in a four-dimensional form with the help
of the unit normal $n_\mu$ to the constant time hypersurfaces,
that is
\begin{equation}\label{ndef}
    n_\mu=\alpha~\delta^0_\mu\qquad
    n^\mu=(-\delta^\mu_0+\delta^\mu_i\beta^i)/\alpha\,.
\end{equation}
We can now decompose the four-dimensional Christoffel symbols in
terms of the standard 3+1 quantities (see Table I). This provides
a convenient way of translating 3+1 conditions like
(\ref{gen_harm}) in terms of four-dimensional objects.

\begin{table}
\centering
  \begin{tabular}{l l}
\hline\noalign{\smallskip}
  ${\Gamma}_{nnn} = 1/\alpha^2\,(\partial_t-\beta^r\partial_r)\,\alpha\;\;\;$ &
  $\;\;\;{\Gamma}_{nnk} = -\partial_k ln\,\alpha$ \\
 \noalign{\smallskip}
  ${\Gamma}_{knn} = 1/\alpha^2\,\gamma_{kj}
  (\partial_t-\beta^r\partial_r)\beta^j + \partial_k ln\,\alpha\;\;\;$ &
  $\;\;\;{\Gamma}_{nij} = -K_{ij}$ \\
\noalign{\smallskip}
  ${\Gamma}_{ijn} = K_{ij}
  -1/\alpha\,\gamma_{ik}\,\partial_j\,\beta^k \;\;\;$ &
 $\;\;\;{\Gamma}_{kij} = ~^{(3)}\Gamma_{kij}$ \\
\noalign{\smallskip}\hline
  \end{tabular}
\smallskip \caption{The 3+1 decomposition of the four-dimensional
connection
  coefficients. The index $n$ is a shorthand for the contraction
  with the unit normal $n_\mu$.}
\label{Gammas}
\end{table}

We can obtain in this way, after an straightforward calculation,
the gauge sources corresponding to the class of slicing conditions
(\ref{gen_harm}), namely
\begin{equation}\label{Z^0}
    H^0 = (1-1/f)~{\Gamma^0}_{\rho\sigma}n^\rho n^\sigma\,.
\end{equation}
We will use now (\ref{harm_gs}) for replacing the quantity $Z^0$
in the Z4 equations. Its evolution equation gets transformed in
this way into a second order evolution equation for the lapse
function $\alpha$, which governs the spacetime slicing. As the
first-order slicing condition (\ref{gen_harm}) has been translated
into a specification of $Z^0$, and allowing for (\ref{Z4div}),
(\ref{gen_harm}) will become a first integral of the second order
evolution system: we can impose it just in the initial data
together with the energy-momentum constraints. This approach is
new in 3+1 formalisms, but a common practice in the harmonic-like
ones.

The same technique can be used for '\,gamma-driver' shift
prescriptions. A first-order reduction of the original
'\,gamma-freezing' condition~\cite{ABDKPST03} is given
by~\cite{MBKC06}
\begin{equation}\label{gamma_driv}
(\partial_t -\beta^k\partial_k)~\beta^i =
\mu~\tilde{\Gamma}^i - \eta ~\beta^i\,,
\end{equation}
where $\tilde{\Gamma}^i$ stands here for the contraction of the
three-dimensional conformal connection, that is
\begin{equation}\label{gamma_tilde}
\tilde{\Gamma}^i \equiv
\gamma^{ij}\,\gamma^{rs}\,(\,\Gamma_{j\,rs} +
\frac{1}{3}~\Gamma_{rsj}\,)\,.
\end{equation}
The corresponding '\,gamma-driver' gauge sources are given by
\begin{eqnarray}\label{Z_i}
    H_i &=&
    (1-\frac{\alpha^2}{\mu})\,{\Gamma}_{i\rho\sigma}n^\rho n^\sigma
    + \frac{1}{3}~{\Gamma}_{\rho\sigma i}\,g^{\rho\sigma}
    \\     \nonumber
&+&(\frac{1}{3}-\frac{\alpha^2}{\mu})\,{\Gamma}_{\rho\sigma
i}n^\rho n^\sigma -\eta/\mu~g_{0i}\,.
\end{eqnarray}
We can use again (\ref{harm_gs}), this time for replacing the
space vector $Z_i$ in the Z4 equations. Its evolution equation get
transformed in this way into a second order evolution equation for
the shift components $\beta^i$, which determine the time lines.
Again, the first-order gamma-driver condition (\ref{gamma_driv})
becomes a first integral of the resulting (second order) shift
evolution equation. At the same time, one gets rid of the
additional variables $Z_i$ (as we did for $Z^0$ with the analogous
replacement, leading to the lapse evolution equation).

\section{Conclusions and outlook}

In summary, we are proposing the action (\ref{action}), which
generalizes the Einstein-Hilbert one. Starting from this action
one gets directly the Z4 field equations, plus the metric
connection condition (which is to be enforced '\,a posteriori' in
our Palatini approach), plus the constraints (\ref{Zis0}) stating
the vanishing of $Z_\mu$. We have shown how a suitable treatment
of these constraints allows working with the Z4 covariant
evolution in the way one usually does in numerical relativity. The
price to pay for this general-covariant approach is that closing
the evolution system requires a separate coordinate gauge
specification. The challenge is then to incorporate the evolution
equations for the gauge-related quantities (lapse and shift) into
the canonical formalism, either via Lagrange
multipliers~\cite{B08,HR10} or by any other means.

We have also presented  an alternative strategy, based in the
'\,gauge sources' approach, which characterizes the generalized
harmonic formalisms. This allows to dispose of the additional
$Z_\mu$ vector field by enforcing at the same time the required
coordinate conditions by means of some generalized gauge sources.
The advantage of this second approach, at the price of getting a
non-covariant evolution system, is that it can allow a direct use
of symplectic integrators, devised to ensure constraint
preservation during numerical evolution (see for instance
ref.~\cite{BGP05}). We have actually identified the gauge sources
corresponding to some standard 3+1 gauge conditions, like the
'puncture gauge' consisting of the '\,1+log' lapse plus the
gamma-driver shift prescriptions. The fact that these popular
gauge conditions can fit into a Lagrangian/Hamiltonian approach,
in the way we have shown, opens the door to new numerical
relativity developments.

{\bf Acknowledgments} The authors are indebted with Luis Lehner
for useful comments and suggestions. This work has been jointly
supported by European Union FEDER funds and the Spanish Ministry
of Science and Education (projects FPA2007-60220 and
CSD2007-00042). C.~Bona-Casas acknowledges the support of the
Spanish Ministry of Science under the FPU program. C.~Palenzuela
acknowledges support from Louisiana State University through the
NSF grant PHY-0803629.

\bibliographystyle{prsty}

\end{document}